\documentclass[journal,transmag]{IEEEtran}
\usepackage{amsmath}
\usepackage{graphicx}
\usepackage{subcaption}
\usepackage{float}
\usepackage{multirow}
\usepackage[table,xcdraw]{xcolor}
\usepackage[noadjust]{cite}

\begin{document}
\title{Optimal Time-Domain Sinusoidal Pulse Width Modulation Technique}
\author{ \IEEEauthorblockN{Siddharth Tyagi and Isaak Mayergoyz, \IEEEmembership{Fellow,~IEEE}}
\IEEEauthorblockA{Department of Electrical and Computer Engineering, University of Maryland, College Park, MD 20742 USA}
\thanks{Corresponding author:  S. Tyagi, email: styagi@umd.edu}}
\maketitle

\begin{abstract}
An optimal time-domain pulse width modulation technique is presented for single-phase and three-phase inverters under the constraint of sinusoidally modulated voltage pulse widths. Harmonic content in currents and voltages is expressed as function of displacement factors, which characterize the placement of the sinusoidally modulated voltage pulses in each sampling subinterval. Implications of symmetries on these displacement factors for three-phase inverters are discussed. Minimization of harmonics is stated as an optimization problem, which is then numerically solved to reveal improvements in harmonic performance. 
\end{abstract}

\newcommand{\dpl}{\Delta t_{\text{pulse}}^{(l)}}
\newcommand{\dpz}{\Delta t_{\text{zero}}^{(l)}}

\section{\textbf{Introduction}}

The pulse width modulation (PWM) technique is widely used to generate AC voltages from a DC voltage source in single-phase and three-phase inverters \cite{holmes2003pulse,patrick2014fundamentals}. The principle of PWM is to generate the inverter voltage as a train of rectangular pulses. The widths of these pulses are modulated to suppress the lower order harmonics at the expense of higher order harmonics. 

Sinusoidal PWM is a technique where the width of the voltage pulse in each switching time-interval is proportional to the value of the desired sinusoidal voltage sampled in the middle of that interval \cite{holmes2003pulse, patrick2014fundamentals}. In this manuscript, an optimal time-domain PWM technique is described for single-phase as well as three-phase inverters when the voltage pulse widths are sinusoidally modulated. This optimal performance is achieved by computing the optimal placement of these sinusoidally modulated pulses in the each switching time-interval.  

This manuscript is organized as follows. Section II describes the optimal PWM technique for single-phase inverters, while Section III deals with three-phase inverters. Section IV presents the numerical results for the optimal PWM technique for both single-phase and three-phase inverters.  
\section{\textbf{Single-Phase Inverters}}

\begin{figure}
	\centering
	\includegraphics[width=1.2\linewidth]{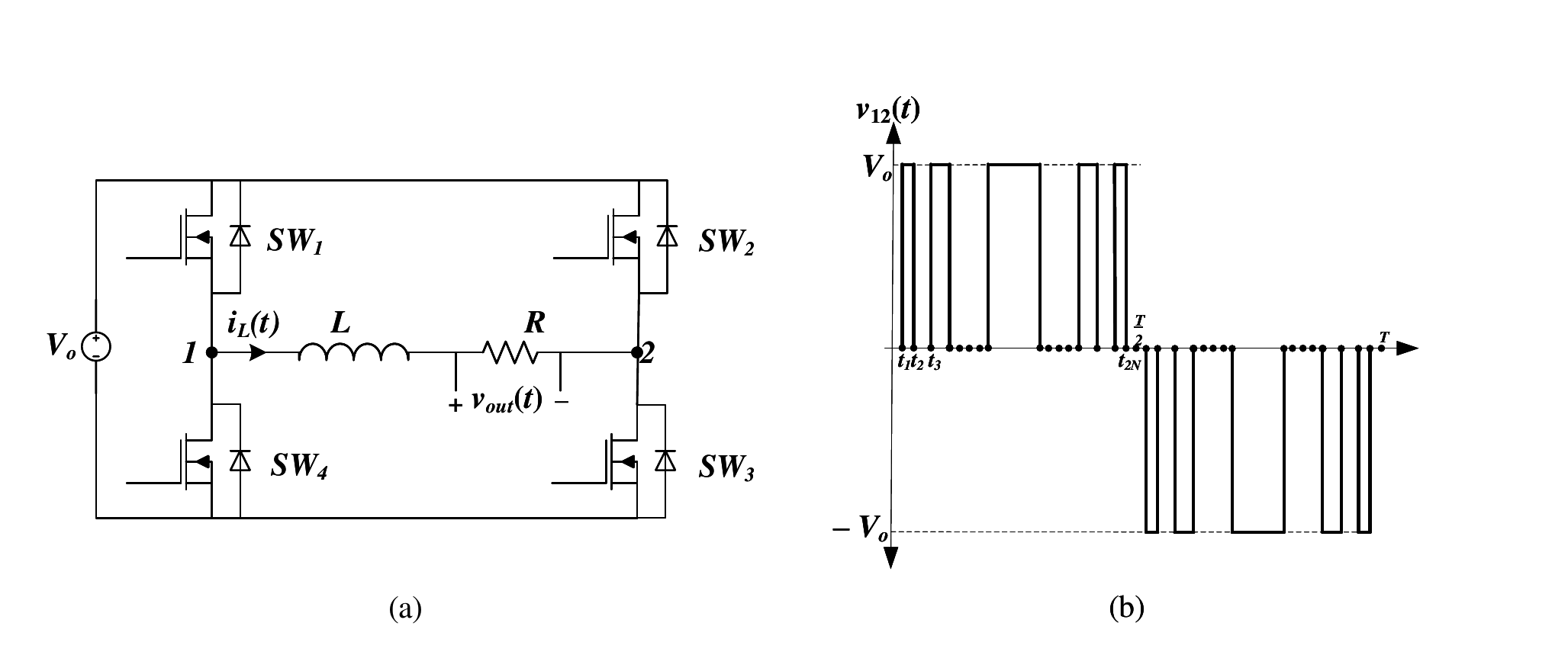}
	\caption{(a) Single-phase H-bridge inverter with $R$-$L$ branch (b) PWM inverter voltage.}
	\label{fig1}
\end{figure}

A single-phase H-Bridge inverter with an $R$-$L$ branch is shown in Fig.~\ref{fig1}(a). The switches $SW_1$, $SW_2$, $SW_3$ and $SW_4$ are used to generate the inverter voltage $v_{12}(t)$ as a train of rectangular pulses of height $\pm V_o$, as shown in Fig.~\ref{fig1}(b). Here, $T$ is the time-period, frequency is $\omega = \frac{2\pi}{T}$, and $N$ is the number of pulses in the half-period $0 < t < T/2$. 

In order to eliminate all even-order harmonics, the voltage $v_{12}(t)$ must satisfy the half-wave symmetry condition:

\begin{equation}
	v_{12}\left( t + \frac{T}{2}\right) = - v_{12}(t), \text{  for all  }  0 < t < \frac{T}{2}.
\end{equation}
The output voltage $v_{12}(t)$ can be characterized by a monotonically increasing sequence of switching time-instants $t_1$, $t_2$, ..., $t_{2N}$, as shown in Fig.~\ref{fig1}(b). Furthermore, we can define: 
\begin{equation}
t_{0} = 0 \text{ and } t_{2N+1} = \frac{T}{2}.
\end{equation}

Using the above switching time-instants, we can derive the analytical expression for the output current $i_L(t)$, which is given by the following formulas \cite{patrick2014fundamentals, tyagi2020optimal}:
\begin{equation}
i_{L}(t)=\begin{cases}
A_{2j+1}e^{-\frac{R}{L}t}, & \text{if $ t_{2j} < t < t_{2j+1} $},\\
A_{2j+2}e^{-\frac{R}{L}t} + \frac{V_o}{R}, & \text{if $ t_{2j+1} < t < t_{2j+2}$},
\end{cases} \label{it}
\end{equation}
where,
\begin{equation} A_1 = -\frac{V_o}{R} \frac {\sum_{j=1}^{2N}{(-1)^j e^{\frac{R}{L}t_j}}}{1 + e^{-\frac{RT}{2L}}} e^{-\frac{RT}{2L}}, \label{A1} \end{equation}  
and
\begin{equation} 
A_j = A_1 + \frac{V_o}{R} \sum_{n=1}^{j-1}{(-1)^n e^{\frac{R}{L}t_n }} \; \;  \text{ for } j=2,3,...,2N+1. \label{Aj} 
\end{equation}

%
%
\subsection{Sinusoidal Pulse Width Modulation}
In PWM, it is customary to modulate the widths of the voltage pulses proportional to the desired output voltage sinusoid. This is achieved as follows. Each half-period $0 < t < T/2$ is subdivided into $N$ equal subintervals. The centers of these subintervals can be specified by the formula:
\begin{equation}
\tau_{sl} = \frac{T}{2N} \left( l - \frac{1}{2}\right) \text{ for } l = 1, 2, ... , N. \label{mid1}
\end{equation}
If the desired inverter voltage $v_{12}^*(t)$ is:
\begin{equation}
	v_{12}^*(t) = V_m \sin \omega t, \label{des1}
\end{equation}
then the width of the $l^{\text{th}}$ pulse $\dpl$  can be obtained as:
\begin{equation}
\Delta t_{\text{pulse}}^{(l)} = \frac{mT}{2N}  \sin \omega \tau_{sl},  \label{dtl1}
\end{equation}
where $m$ is the modulation index defined as: 
\begin{equation}
	m = \frac{V_m}{V_o}. \label{modi}
\end{equation}

This implies that for sinusoidal pulse width modulation, the switching time-instants satisfy the following condition:
\begin{equation}
t_{2l} - t_{2l - 1} =\dpl \text{ for } l = 1, 2, ..., N,
\end{equation}
where $\dpl$ is defined by equation (\ref{dtl1}). The above constraint does not specify the switching time-instants uniquely. We have the additional freedom to choose the position of these pulses within each subinterval of length $T/2N$. To explore this, we define the time $\dpz$ as:
\begin{equation}
 \dpz = \frac{T}{2N} - \dpl \label{dlz1}.
\end{equation} 
Evidently, $\dpz$ is the time duration within the $l^{\text{th}}$ subinterval for which the voltage $v_{12}(t)$ is $0$.
	\begin{figure}[h!]
	\centering
	\includegraphics[trim={0 0.5cm 0 1.2cm},width=0.85\columnwidth]{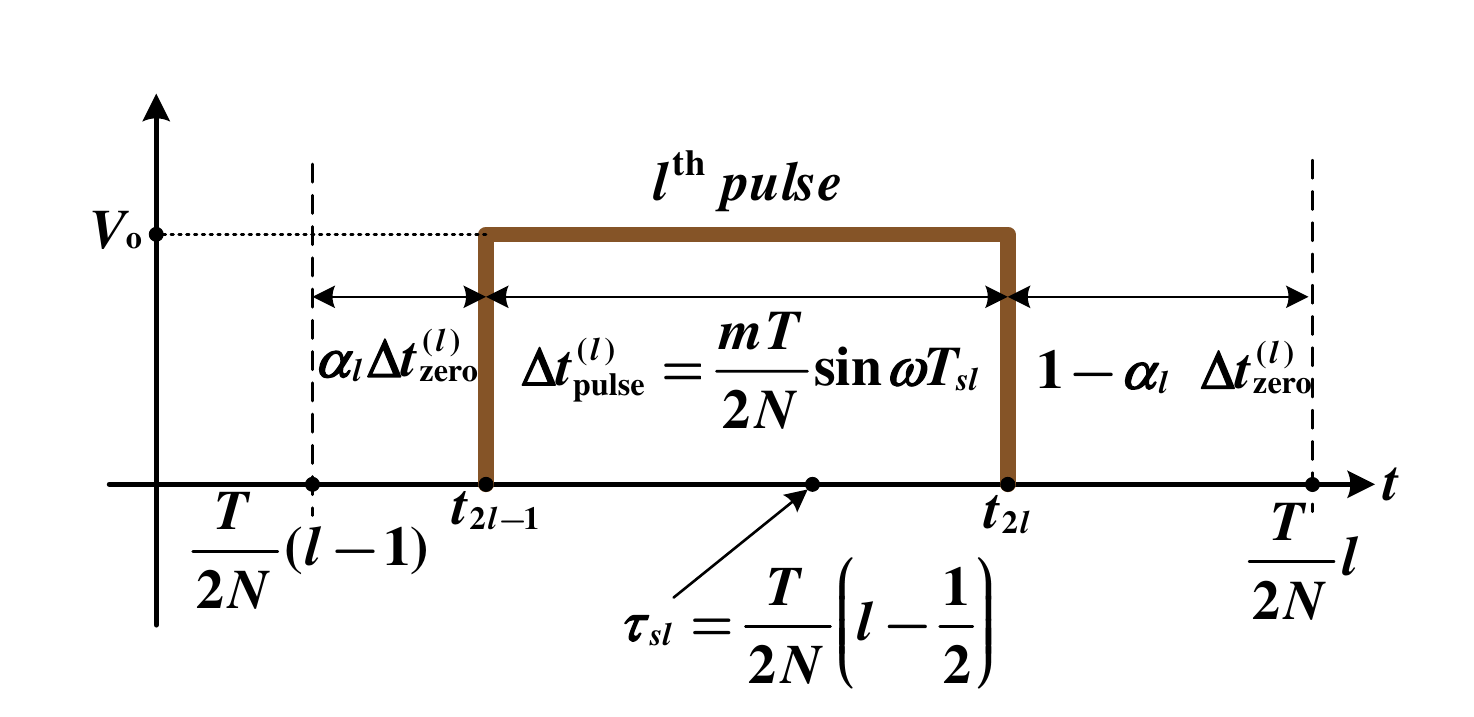}
	\caption{Time-instants for sinusoidally modulated $l^{\text{th}}$ pulse}
	\label{lpulse1}
\end{figure}

 As shown in Fig.~\ref{lpulse1}, we define the \textit{displacement factor} $\alpha_l$ such that the time between the beginning of the $l^{\text{th}}$ subinterval and the rising edge of the $l^{\text{th}}$ voltage pulse is $\alpha_l \dpz$. Clearly,
\begin{equation}
	0 \leq \alpha_l \leq 1 \text{ for each } l = 1, 2, ... , N. \label{aeq}
\end{equation}
Thus, in general the switching time-instants for sinusoidally modulated pulses can be expressed as:
\begin{align}
t_{2l - 1} &= \frac{T}{2N} (l - 1) + \alpha_l \dpz, \label{tl1} \\
t_{2l} & = t_{2l - 1} + \dpl. \label{tl2}
\end{align}
It is evident from equations (\ref{dtl1}), (\ref{dlz1}), (\ref{tl1}) and (\ref{tl2}) that for a given modulation index $m$, time-period $T$ and pulse-number $N$, single-phase sinusoidally modulated voltage pulses can be completely characterized by $N$ displacement factors $\alpha_1$, $\alpha_2$, $...$, $\alpha_N$. Consequently, according to formulas (\ref{it}), (\ref{A1}) and (\ref{Aj}) the output current $i_L(t)$ can also be expressed as a function of these  displacement factors. That is, for sinusoidally modulated voltage pulses,
\begin{align}
	i_{L}(t) &= i_{L}(t, \alpha_1, \alpha_2, ... , \alpha_N),
\end{align}
For the conventional sinusoidal PWM, the voltage pulses are placed in the center of the corresponding time subintervals, and hence :
\begin{equation}
	\alpha_1 = \alpha_2 = ... = \alpha_N = 0.5 \text{ for conventional PWM}.
\end{equation}

\subsection{Optimal PWM with sinusoidal pulse widths }
The optimization of the output current harmonics can now be performed by minimizing the following function of the displacement factors $\alpha_1$, $\alpha_2 $, $ ...$, $\alpha_N$:
\begin{align}
&E_{2}(\alpha_1, \alpha_2, ... , \alpha_N) = \nonumber \\ &\int_{0}^{T/2} \left[i_L(t, \alpha_1, \alpha_2, ... , \alpha_N) - I_m \sin\left( \omega t - \phi \right) \right]^2 dt,
\end{align}
where 
\begin{align}
I_m &= \frac{V_m}{\sqrt{R^2 + (\omega L)^2}},\\
&\tan \phi = \frac{\omega L}{R}.
\end{align}
The function $E_2$ is the $L_2$-norm of the error between the actual output current and its desired value. The optimization of this function is performed over the convex region \cite{nesterov2013introductory} which is a cube defined by the inequalities (\ref{aeq}). For PWM the following quarter-wave symmetry condition is usually satisfied:
\begin{equation} v_{12}(t) =  v_{12}\bigg( \frac{T}{2} - t \bigg).  \label{ss3} \end{equation}
From the above equation, the following linear relation between the displacement factors can be derived:
\begin{equation}
	\alpha_l + \alpha_{N+1-l} =1. \label{acons2}
\end{equation}
Hence, the number of independent displacement factors is reduced from $N$ to $(N-1)/2$. As a result, the number of independent variables involved in the optimization is reduced as well. Furthermore, the optimization can be modified by introducing specific constraints which result in the selective harmonic elimination \cite{holmes2003pulse,mayergoyz2018optimal}. The numerical results of the optimization are presented in the last section. 

\section{\textbf{Three-Phase Inverters}}
\begin{figure}[h]
	\centering
	\includegraphics[trim = {0, 0, 0, 0.5cm}, width=0.7\columnwidth]{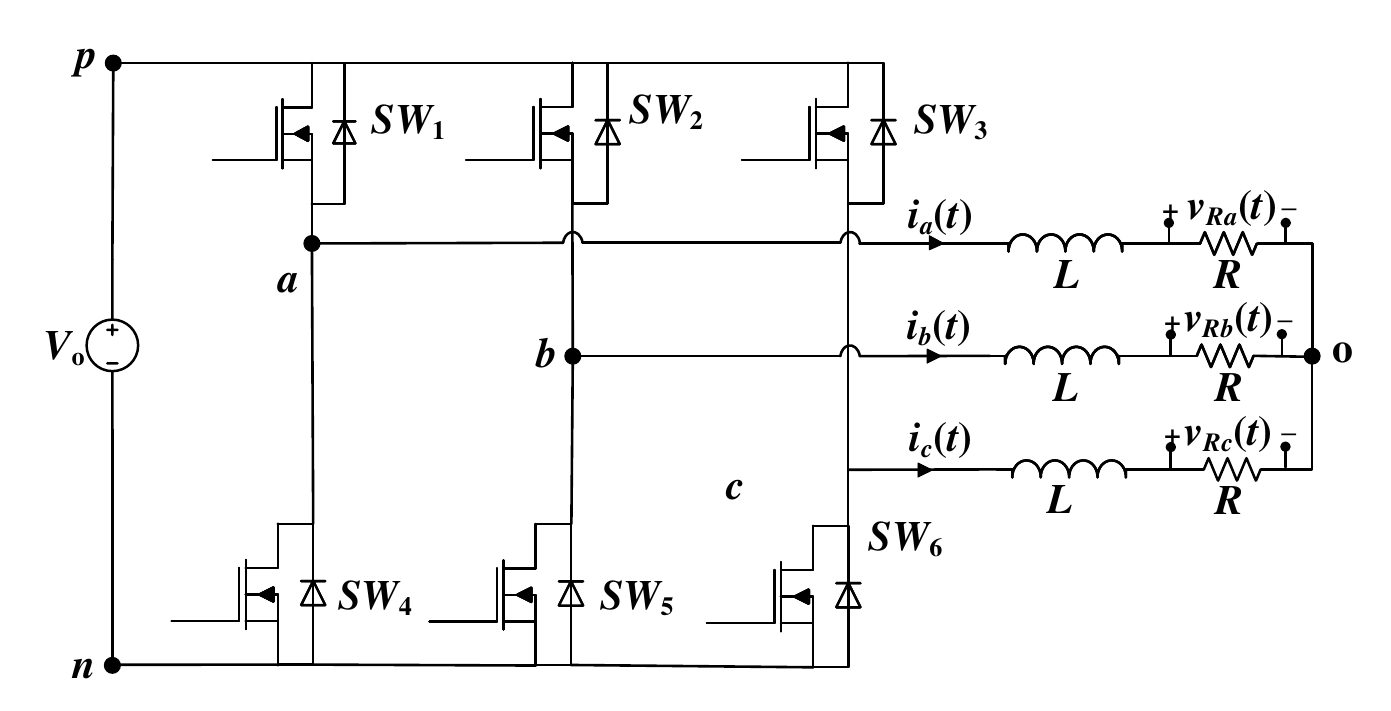}
	\caption{3-phase H-Bridge Inverter with $R$-$L$ Branch}
	\label{3pinv}
\end{figure}

A three-phase H-Bridge inverter with an $R$-$L$ branch is shown in Fig.~\ref{3pinv}. The three line-voltages $v_{ab}(t), v_{bc}(t)$ and $v_{ca}(t)$ are periodic trains of rectangular pulses with period $T$ and frequency $\omega = \frac{2\pi}{T}$, such that each half-period has $N$ pulses, similar to the single-phase inverter voltage (see Fig.~\ref{fig1}(b)). 

These voltages satisfy the following symmetry conditions \cite{tyagi2020optimal}:

\textbf{S1. Translational Symmetry}: The three-phase line-voltages are $\frac{T}{3}$-time shifted  versions of each other. That is,
\begin{equation} v_{ab}(t) = v_{bc}\bigg(t + \frac{T}{3} \bigg) = v_{ca}\bigg(t - \frac{T}{3} \bigg). \label{S1} \end{equation}
The above condition leads to the elimination of all harmonic multiples of three.  

\textbf{S2. Half-Wave Symmetry}: This means:
\begin{equation} v_{ab}(t) = - v_{ab}\bigg(t + \frac{T}{2} \bigg).  \label{S2} \end{equation}

\textbf{S3. Quarter-wave symmetry}:    
\begin{equation} v_{ab}(t) =  v_{ab}\bigg( \frac{T}{2} - t \bigg).  \label{S3} \end{equation}

Half-wave symmetry in these voltages implies that they can be completely characterized by switching time-instants $t_1$, $t_2, ...,$ $t_{2N}$ in the half period $ 0 < t < T/2$. Using per-phase analysis and the above symmetries, the analytical expression for the phase current $i_a(t)$ in terms of the switching time-instants can be derived as \cite{tyagi2020optimal}: 
	\begin{equation}
i_a(t) = \frac{1}{3}\bigg[i_{ab}(t,t_1, t_2, ... , t_{2N}) - i_{ab}\bigg(t + \frac{T}{3}, t_1, t_2, ... , t_{2N} \bigg)\bigg],
\label{ia} 
\end{equation}
where $i_{ab}(t)$ has the same analytical form as $i_L(t)$ for the single-phase inverter presented in equations (\ref{it})-(\ref{Aj}).
From equation (\ref{ia}), we conclude that 
\begin{equation}
i_a(t) = i_a(t, t_1, t_2, ... , t_{2N}). \label{iafunc}
\end{equation}

\begin{figure}[h]
	\includegraphics[trim={0 0.5cm 0 1cm}, scale=0.6]{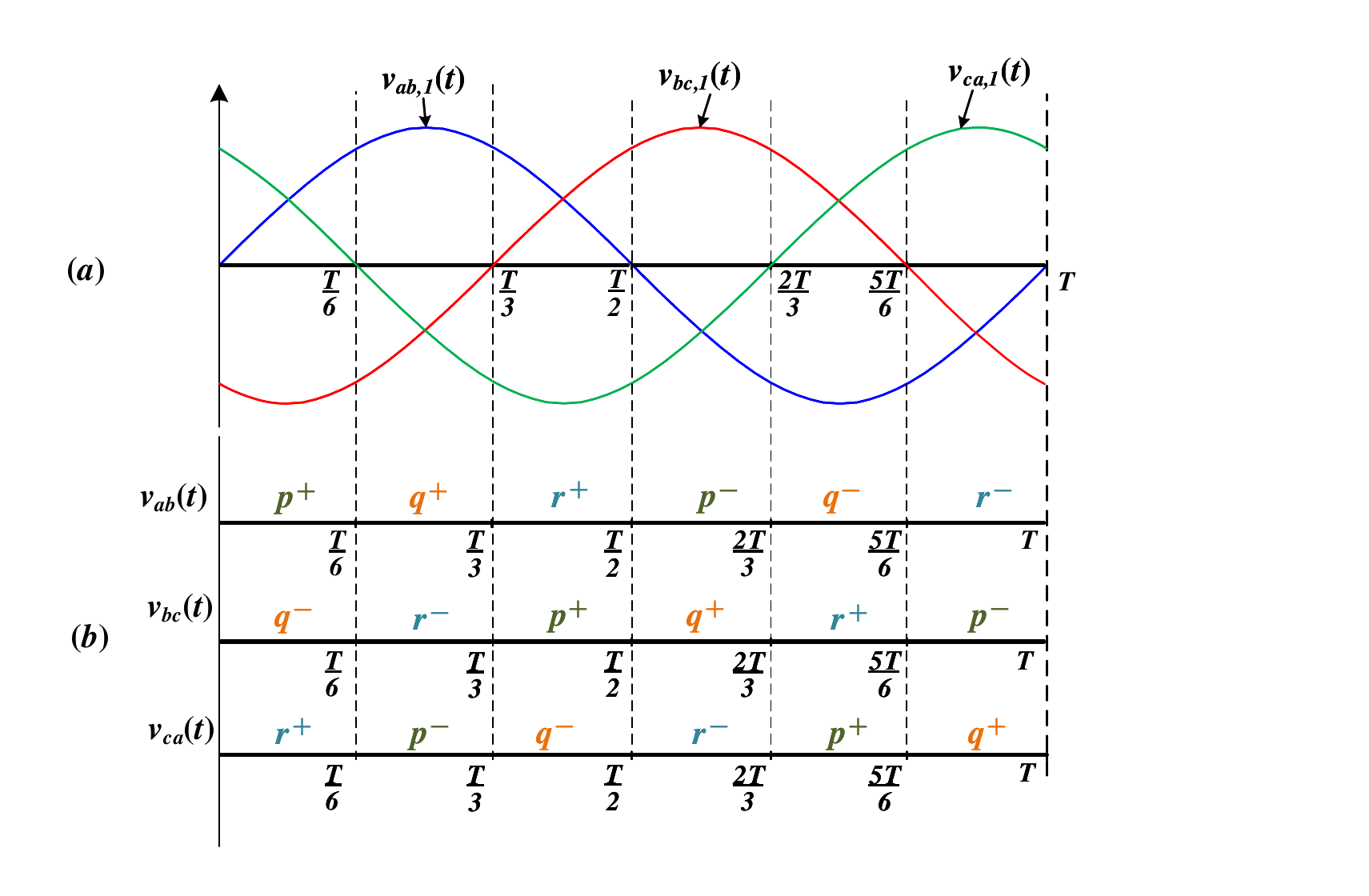}
	\caption{Structure of 3-phase line-voltages}
	\label{pqrdiag}
\end{figure}

The above symmetry conditions as well as the Kirchhoff voltage law (KVL) and the condition that the transistors only in one leg of the inverter can be switched at any instant of time, have implications on the structure of line-voltages and impose algebraic constrains on the switching time-instants for three-phase line-voltages \cite{tyagi2020optimal}.  These can be understood through Fig.~\ref{pqrdiag}. Here, for each line-voltage, pulses within an interval of length $T/6$ form a pulse group, which can be labeled as $p^+$, $q^+$, $r^+$, $p^-$, $q^-$ and $r^-$.
It can be shown \cite{tyagi2020optimal} that there are equal number of pulses $P$ in each interval of length $T/6$ for the line-voltages, i.e. $N = 3P$. Furthermore,  $P$ should be an odd number to have a pulse at $t = T/4$ for $v_{ab}(t)$, where this voltage is maximum. 
 The switching time-instants for pulses in the $p$-group ($t_k$ for all $k = 1, 2, .. , 2P$) completely determine the switching time-instants for pulses in the $r$-group using quarter-wave symmetry:
\begin{equation}
	t_{k} + t_{6P+1-k} = \frac{T}{2}, \text{ for all } k = 1, 2, ..., 2P. \label{pr}
\end{equation}
Here, $t_{6P+1-k}$ are switching time-instants for pulses in the $r$ group.
KVL implies that pulses in the $q$-group are sums of pulses in the $p$ and $r$ groups of the opposite polarity. The condition that transistors in only one leg of the inverter can switch at any instant of time implies that the order in which $p$ and $r$-group pulses sum up to produce $q$-group pulses alternates.  Using this, algebraic relations can be derived to obtain switching time-instants for pulses in the $q$-group from those in $p$-group. 
Thus, the line voltages can be completely characterized by the time-instants $t_1$, $t_2$, $...$, $t_{2P}$ in the interval $ 0 < t < T/6$ corresponding to the $p^+$-group of the line-voltage $v_{ab}(t)$. Moreover, there are additional algebraic constraints these time-instants. The details of the above discussion can be found in \cite{tyagi2020optimal}.

\subsection{Sinusoidal Pulse Width Modulation} 

\begin{figure}[h]
	\includegraphics[trim={0 1cm 0 1.2cm},scale=0.50]{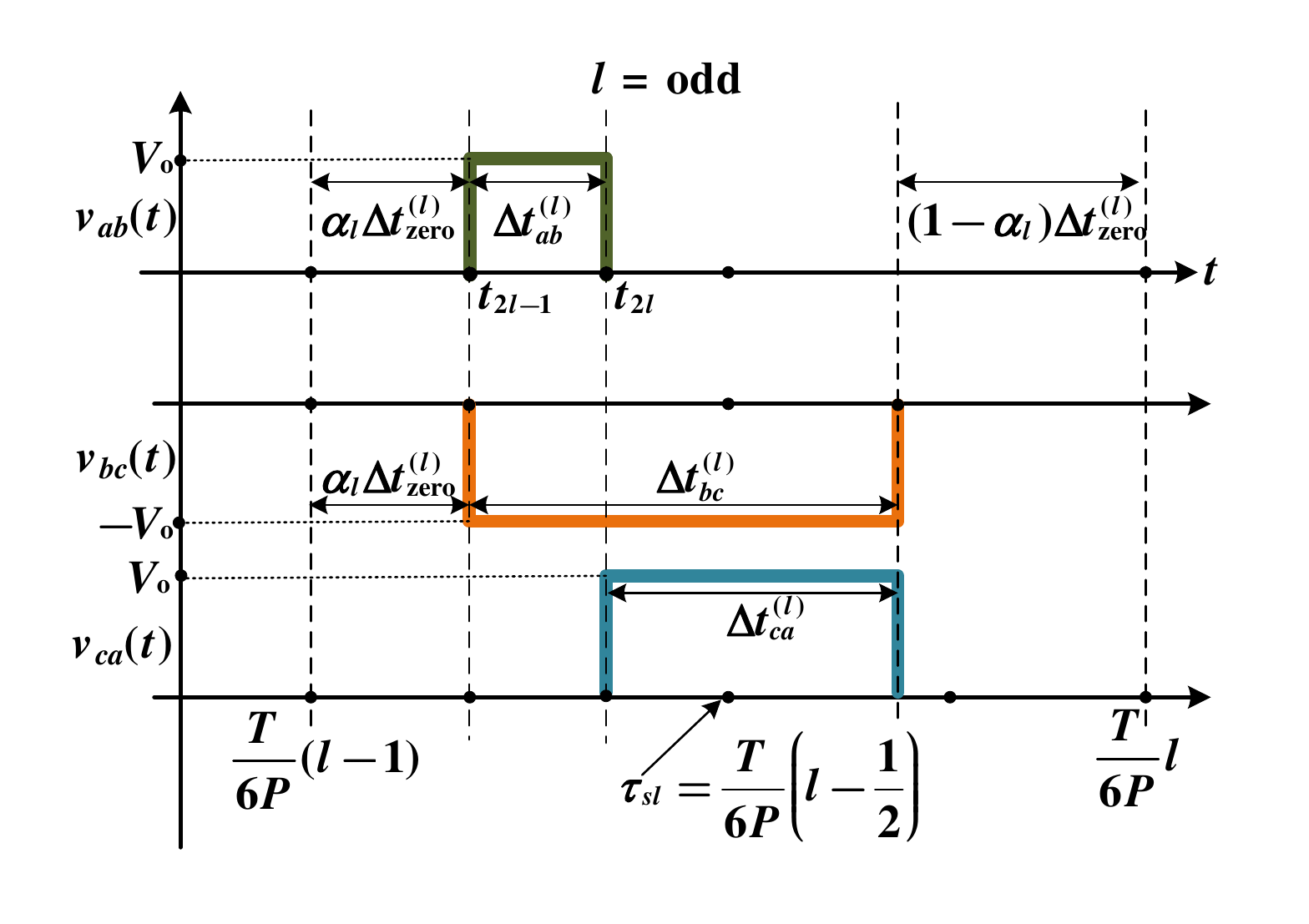}
	\caption{Line-voltage pulses for the case when $l$ is odd.}
	\label{3lodd}
\end{figure}

Consider the interval $0 < t < T/6$, where pulses for the line-voltages $v_{ab}(t)$, $v_{bc}(t)$ and $v_{ca}(t)$ belong to the $p^+,q^-$ and $r^+$ groups, respectively. This interval is divided into $P$ subintervals to have $P$ pulses for each line-voltage. Consider the $l^{\text{th}}$ pulse in this interval for the three line-voltages. 
The widths of these pulses are sinusoidally modulated to obtain the desired line voltages $v_{ab}^*(t)$, $v_{ab}^*(t)$ and $v_{ab}^*(t)$ which have a peak value $V_m$ and are balanced, positive sequence three phase voltages \cite{patrick2014fundamentals}. Again, defining the modulation index $m$ using equation (\ref{modi}), the widths of the line-voltage pulses can be obtained as: 
 \begin{align}
\Delta t_{ab}^{(l)} &= \frac{mT}{6P}\sin \omega \tau_{sl} \label{dtab}, \\
\Delta t_{bc}^{(l)} & = -\frac{mT}{6P} \sin\bigg(\omega \tau_{sl}- \frac{2\pi}{3}\bigg), \label{dtbc} \\
\Delta t_{ca}^{(l)} &= \frac{mT}{6P} \sin\bigg(\omega \tau_{sl} + \frac{2\pi}{3}\bigg), \label{dtca}
\end{align}
where
\begin{equation}
	\tau_{sl} = \frac{T}{6P} \left( l - \frac{1}{2}\right).
\end{equation}
The duration $\dpz$ for which all the three line-voltages are zero can be obtained as:
\begin{equation}
	\dpz = \frac{T}{6P} - \Delta t_{bc}^{(l)}. \label{dtz}
\end{equation}
As in the case for the single-phase inverter, there is an additional freedom to place these pulses within each subinterval.  The structure of the three line-voltages for the case when $l$ is odd is shown in Fig.~\ref{3lodd}. For this case, $\alpha_l$ is the \textit{displacement factor} such that the time between the beginning of the $l^{\text{th}}$ interval and the rising edge of the pulse for $v_{ab}(t)$ is $\alpha_l \dpz$. In this case, the time-instants $t_{2l-1}$ and $t_{2l}$ can be obtained as:
\begin{align}
	t_{2l-1} &= \frac{T}{6P}\left( l - \frac{1}{2}\right) + \alpha_l \dpz, \label{tro} \\
	t_{2l} &= t_{2l-1} + \Delta t_{ab}^{(l)}. \label{tfo}
\end{align}
 
\begin{figure}[h]
	\includegraphics[trim={0 1cm 0 1.2cm}, scale=0.50]{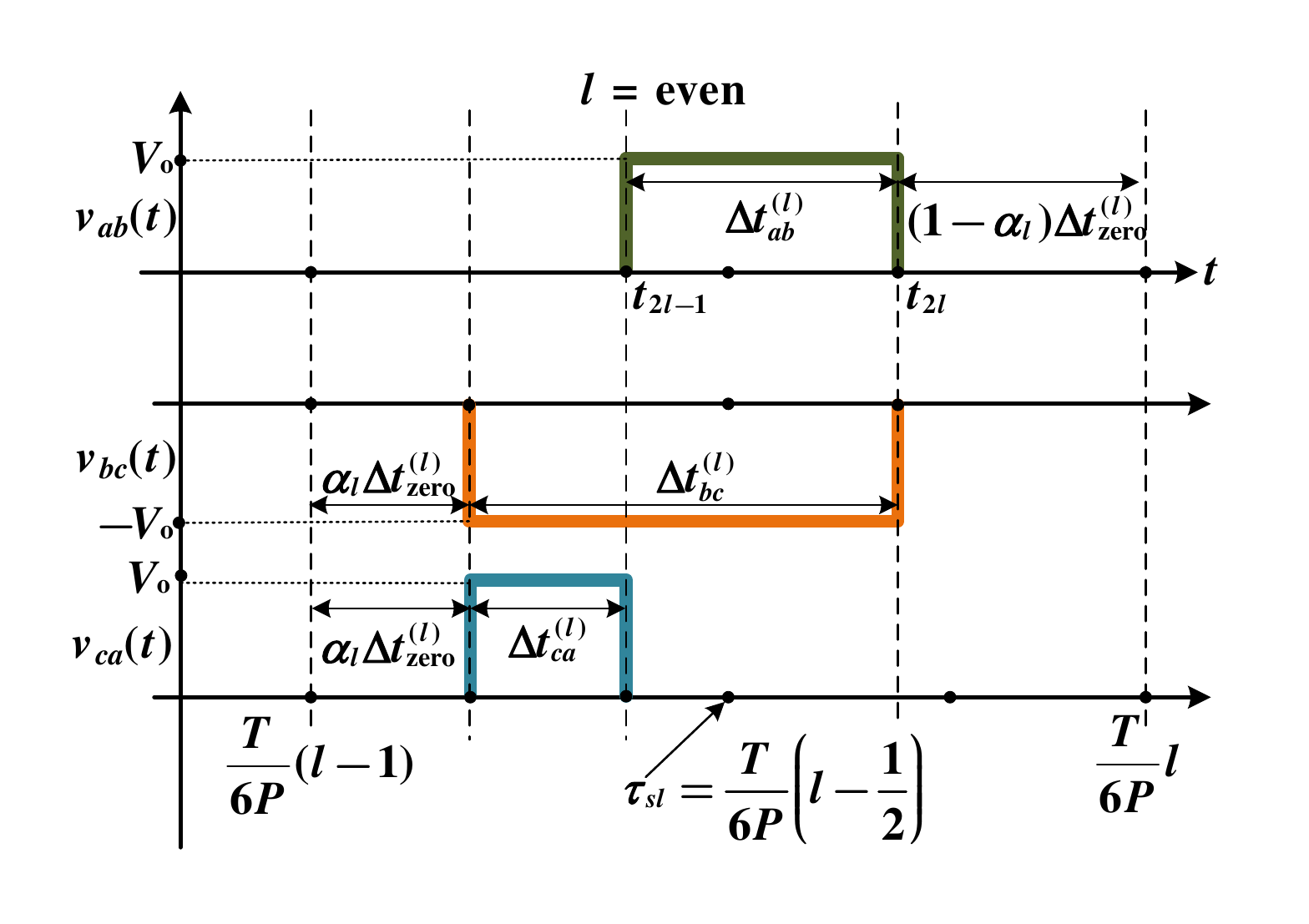}
	\caption{Line-voltage pulses for the case when $l$ is even.}
	\label{3leven}
\end{figure}

Similarly, when $l$ is even, we can derive (see Fig.~\ref{3leven})
\begin{align}
t_{2l-1} &= \frac{T}{6P}(l - 1) + \alpha_l \dpz+ \Delta t_{ca}^{(l)} ,\label{tre} \\
t_{2l} &= t_{2l-1} + \Delta t_{ab}^{(l)}. \label{tfe}
\end{align}
It is evident from equations (\ref{dtab})- (\ref{dtca})  and (\ref{tro})-(\ref{tfe}) that for given modulation index $m$, time-period $T$ and number of pulses in each group $P$, three-phase sinusoidally modulated voltage pulses can be completely characterized by $P$ displacement factors $\alpha_1$, $\alpha_2$, $...$, $\alpha_P$.

It turns out that not all $\alpha_l$ are independent. Algebraic relations exist between these displacement factors, which can be obtained as follows. It is clear from Fig.~\ref{3lodd} that when $l$ is odd, the falling edge of the $v_{ab}(t)$ $p$-pulse corresponds to the rising edge of the $v_{ca}(t)$ $r$-pulse. This implies:
\begin{equation}
	t_{2l} = \frac{T}{3} - t_{4P+2l-1}.
\end{equation}
But, from equation (\ref{pr}), if we use $k = P+1-l$, we obtain:
\begin{equation}
 	t_{4P+2l-1}= \frac{T}{2}-t_{2(P+1-l)}.  
	\end{equation}
From the above two, we deduce 
\begin{equation}
	t_{2l} + t_{2(P+1-l)} = \frac{T}{6}.
\end{equation}
Note that if $l$ is odd, $P+1-l$ is also odd. Hence, $t_{2(P+1-l)}$ can be obtained from equation (\ref{tfo}) by putting $l = P+1-l$. In this way, we can derive the following:
\begin{equation}
	\dpz \alpha_l + \Delta t_{\text{zero}}^{(P+1-l)} \alpha_{P+1-l}  = \frac{T}{6P} - \Delta t_{ab}^{(l)} - \Delta t_{ab}^{(P+1-l)}.
\end{equation}

By using (\ref{dtab}) , (\ref{dtz}) and trigonometric identities, we can greatly simplify the above equation to obtain:
\begin{equation}
	\alpha_l + \alpha_{P+1-l} = 1. \label{acons}
\end{equation}
It turns out that the relation (\ref{acons}) is also true when $l=\text{even}$. Thus, the number of independent displacement factors which characterized three-phase sinusoidal PWM are further reduced from $P$ to $(P-1)/2$.

The above presented analysis of sinusoidal PWM for three-phase inverters can be understood as a time-domain interpretation of the space-vector PWM (SVPWM) technique \cite{holmes2003pulse, van1988analysis,ogasawara1990novel}. Indeed, each interval of length $T/6$ shown in Fig.~\ref{pqrdiag} can be associated with a specific sector of  SVPWM technique. The expressions (\ref{dtab}), (\ref{dtbc}) and (\ref{dtca}) for sinusoidal pulse-widths can be also derived for the SVPWM technique.  The time duration $\dpz$ in formula (\ref{dtz}) gives the total duration of the zero space-vectors. Furthermore, the alternation of $p$-group and $r$-group pulses as they sum up to produce the $q$-group pulses is also a characteristic feature of SVPWM. For conventional SVPWM, the two zero space-vectors have equal durations in each switching interval, and hence:
\begin{equation}
	\alpha_1 = \alpha_2 = ... = \alpha_P = 0.5.
\end{equation}
\subsection{Optimal Three-Phase PWM with Sinusoidal Widths}

From the previous subsection, we have deduced that for sinusoidal pulse widths, the switching time-instants which define PWM line-voltages can be obtained from the displacement factors $\alpha_1, \alpha_2, ..$. for pulses in the interval $0 < t < T/6$. This implies that the current $i_a(t)$ from equation (\ref{iafunc}) can be expressed as:
\begin{equation}
	i_a(t) = i_a(t, \alpha_1, \alpha_2, ...). 
\end{equation}
The harmonics in this current can be minimized by minimizing the following function, 
\begin{align}
&\tilde{E}_{2}(\alpha_1, \alpha_2, ... ) = \nonumber \\ &\int_{0}^{T/2} \left[i_L(t, \alpha_1, \alpha_2, ... ) - \tilde{I}_m \sin\left( \omega t - \tilde{\phi} \right) \right]^2 dt,
\end{align}
where the variable $\alpha_1, \alpha_2$,... satisfy the condition (\ref{aeq}) as well as the constraints (\ref{acons}), and:
\begin{align}
	\tilde{I}_m = \frac{V_m}{\sqrt{3}.\sqrt{R^2 + (\omega L)^2}}, \\
	\tilde{\phi} = \frac{\pi}{6} + \arctan{\frac{\omega L}{R}}.
\end{align}

\section{\textbf{Numerical Results}}
The described optimization techniques for single-phase and three-phase inverters have been numerically implemented by using the Interior point method \cite{byrd1999interior, nocedal2006numerical} which is provided in MATLAB. The values parameters were chosen as follows: $V_o = 300$ V, frequency $f = 60$ Hz, and $I_m = 10$ A. The optimization was performed for different values of $m$, $L$ and $N$ (for single-phase) and $P$ (for three-phase). 
Sample values of optimal displacement factors are shown in Table \ref{taba}. It can be observed that for single-phase inverters these displacement factors are monotonically decreasing until they reach 0.5 (=$\alpha_{(N+1)/2}$, from (\ref{acons2})). For three-phase inverters, the displacement factors first monotonically decrease before increasing towards 0.5 (= $\alpha_{(P+1)/2}$, from (\ref{acons})). These trends are consistently observed for different values of $N$ (or $P$), $m$ and $L$.
\begin{table}[]
	\caption{For $N=11$ and $P=11$, there are $5$ independent $\alpha$ parameters. Values of $\alpha$'s for $L = 100 \mu$H and $m=0.9$.}
	\begin{tabular}{l|c|c|c|c|c|}
		\cline{2-6}
		& $\alpha_1$                          & $\alpha_2$                         & $\alpha_3$                         & $\alpha_4$                          & $\alpha_5$                         \\ \hline
		\multicolumn{1}{|l|}{Single-Phase} & 0.9567                      & 0.8621                      & 0.8347                      & 0.7837                      & 0.6410                      \\
		\multicolumn{1}{|l|}{Three-Phase}  & \multicolumn{1}{l|}{0.9776} & \multicolumn{1}{l|}{0.7652} & \multicolumn{1}{l|}{0.4322} & \multicolumn{1}{l|}{0.2231} & \multicolumn{1}{l|}{0.4457} \\ \hline
	\end{tabular}
\label{taba}
\end{table}

Next, optimal and conventional PWM methods are compared. The total harmonic distortion (THD) \cite{holmes2003pulse, tyagi2020optimal} in the output currents is an important measure of their harmonic content. The THD for conventional (THD$_{\text{conv}}$) and optimal (THD$_{\text{opt}}$) PWM is measured, and the percentage improvement in THD is computed as:
\begin{equation}
\%	\text{improvement}  = \frac{\text{THD}_{\text{conv}} - \text{THD}_{\text{opt}} }{\text{THD}_{\text{conv}}}.
\end{equation} 
 
 First, the impact of modulation index $m$ on conventional and optimal THD values is analyzed (see Table \ref{Tab1}). For both single and three-phase inverters, it is found that a higher $m$ results in a lower THD for the same pulse number $N$ (or $P$) and $L$. Furthermore, the percentage improvement in THD is also found to be higher for greater values of $m$. This leads us to conclude that operating with a higher modulation index is better for the harmonic performance of inverters. This might also help reduce iron-losses in the core of the inverters which depend inversely on the modulation index \cite{boglietti1993effects}.
\begin{table}[h!]
	\centering
	\caption{Values of  conventional and optimal THD in single-phase inverters when $N=11$, $L = 100 \mu$H, for various modulation index $m$.}
	\begin{tabular}{|c|c|c|c|}
		\hline
		\multirow{2}{*}{$m$} & \multicolumn{2}{c|}{THD (in \%)} & \multirow{2}{*}{\% improvement} \\ \cline{2-3}
		& conventional  & optimal  &                                 \\ \hline
		0.95               & 36.15         & 30.49    & 15.66                           \\
		0.90               & 39.23         & 33.49    & 14.63                           \\
		0.85               & 42.34         & 36.55    & 13.67                           \\
		0.80               & 45.55         & 39.53    & 12.81                           \\
		0.75               & 48.05         & 42.32    & 11.92                           \\ \hline
	\end{tabular}
\label{Tab1}
\end{table}

Next, the effect of the value of inductance on THD is presented. The values of THD after optimization suggest that better harmonic performance is achievable even with lower inductance by choosing appropriate displacement factors. For instance, from Table \ref{tab2}, the optimal THD for $L = 25 \mu$H is greater than the conventional THD for the (twice in value) inductance of $50 \mu$H. This clearly reveals the efficacy of the developed optimization technique for low inductance values.  
  
\begin{table}[h!]
	\centering
	\caption{Improvement in THD with optimization when $m = 0.95$, $N = 11$, for different values of inductance $L$}
	\begin{tabular}{|c|c|c|c|}
		\hline
		& \multicolumn{2}{c|}{THD (in \%)}                                    &                                  \\ \cline{2-3}
		\multirow{-2}{*}{\begin{tabular}[c]{@{}c@{}}$L$\\ (in $\mu$H)\end{tabular}} & conventional                 & optimal                      & \multirow{-2}{*}{\% improvement} \\ \hline
		125                                                                   & 35.96                        & 29.16                        & 18.91                            \\
		100                                                                   & 39.76                        & 32.53                        & 18.18                            \\
		75                                                                    & {44.03} & {36.49} & {17.12}     \\
		50                                                                    & {48.79} & {41.02} & {15.92}     \\
		25                                                                    & {54.04} & {46.26} & {14.39}     \\ \hline
	\end{tabular}
\label{tab2}
\end{table}

Finally, the effect of pulse number for different values of $N$ (for single-phase inverters) and $P$ (for three-phase inverters) is shown in Tables \ref{tab3} and \ref{tab4}. It is evident that by using optimal displacement factors, better THD can be achieved even with lower pulse numbers (lower switching frequencies). For instance, for the single-phase inverter, the optimal THD for $N=7$ is lesser than the the conventional THD for $N=13$, and is comparable to the THD for $N = 15$ (almost twice the switching frequency). Moreover, the percentage improvement in THD is also greater for lower values of $N$ (or $P$). This highlights the importance of choosing optimal displacement factors when the number of pulses is low. 
\begin{table}[h!]
	\centering
	\caption{Values of THD for single-phase inverters, for different values of $N$, when $m = 0.95$, $L = 100 \mu$H}
	\begin{tabular}{|c|c|c|c|}
		\hline
		\multirow{2}{*}{N} & \multicolumn{2}{c|}{THD (in \%)} & \multirow{2}{*}{\% improvement} \\ \cline{2-3}
		& conventional      & optimal      &                                 \\ \hline
		15                 & 35.58             & 31.40        & 11.75                           \\
		13                 & 38.10             & 33.13        & 13.04                           \\
		11                 & 40.88             & 34.83        & 14.80                           \\
		9                  & 44.03             & 36.49        & 17.12                           \\
		7                  & 47.86             & 37.96        & 20.69                           \\ \hline
	\end{tabular}
\label{tab3}
\end{table}

\begin{table}[h!]
	\centering
	\caption{Values of THD for three-phase inverters, for different values of $P$, when $m = 0.95$, $L = 50 \mu$H.}
	\begin{tabular}{|c|c|c|c|}
		\hline
		\multirow{2}{*}{P} & \multicolumn{2}{c|}{THD (in \%)} & \multirow{2}{*}{\% improvement} \\ \cline{2-3}
		& conventional      & optimal      &                                 \\ \hline
		15                 & 29.71             & 26.36        & 11.26                           \\
		13                 & 31.96             & 28.64        & 10.38                           \\
		11                 & 34.45             & 31.33        & 9.03                            \\
		9                  & 37.22             & 34.04        & 8.55                            \\ \hline
	\end{tabular}
\label{tab4}
\end{table}

\bibliographystyle{ieeetran}
\bibliography{Referen}
\end{document}